\documentclass[
  ,final            
  ]
  {aipproc}

\usepackage{natbib}

\layoutstyle{6x9}

\newcommand\apj{ApJ}%
\newcommand\apjs{ApJS}%
\newcommand\aap{A\&A}%
\begin{document}

\title{The GALEX View of Supernova Hosts}

\classification{97.60.Bw}
\keywords      {galaxies: evolution -- supernovae: general}

\author{James~D.~Neill}{
  address={California Institute of Technology, 1200 E. California Blvd., Pasadena, CA, 91214, USA}
}

\author{Mark~Sullivan}{
  address={University of Oxford, Denys Wilkinson Building,
  Keble Road, Oxford, OX1 3RH, UK}
}

\author{Mark Seibert}{
address={The Observatories of the Carnegie Institute of Washington, 813 Santa Barbara Street, Pasadena, CA, 91101, USA}
}

\begin{abstract} 
	
	We exploit the accumulating, high-quality, multi-wavelength imaging
	data of nearby supernova (SN) hosts to explore the relationship
	between SN production and host galaxy evolution.  The Galaxy
	Evolution Explorer ({\it GALEX}, \citep{Martin:05:L1}) provides
	ultraviolet (UV) imaging in two bands, complementing data in the
	optical and infra-red (IR).  We compare host properties, derived
	from spectral energy distribution (SED) fitting, with nearby,
	well-observed SN~Ia light curve properties.  We also explore where
	the hosts of different types of SNe fall relative to the red and
	blue sequences on the galaxy UV-optical color-magnitude diagram
	(CMD, \citep{Wyder:07:293}).  We conclude that further exploration
	and larger samples will provide useful results for constraining the
	progenitors of SNe.

\end{abstract}

\maketitle


\section{Introduction}

The diagnostic power of {\it GALEX} imaging for galaxy evolution is
revealed in a plot of simple stellar population age versus wavelength shown
at the top of Figure~1 in \citet{Martin:05:L1}. This plot covers the UV to
the IR and demonstrates that the UV is most sensitive to star formation on
time scales of $\sim$10$^8$ yr.  On the same figure, we see the flux of
stellar populations with differing Scalo b ($\equiv$
\.M/$\langle$\.M$\rangle$, where \.M is the star formation rate in
M$_{\odot}$ yr$^{-1}$) plotted against wavelength.  The UV shows more
discriminating power than other wavelengths.  {\it GALEX} imaging is
sensitive over at least five orders of magnitude of Scalo b, from b $= 10$
down to b $= 10^{-3}$.  Another illustration of the leverage of UV imaging
on galaxy evolution is shown in the galaxy CMD in Figure~9 of
\citet{Wyder:07:293}.  Plotting galaxy near-UV minus r-band (NUV-r) color
versus absolute r-band magnitude (M$_r$) clearly separates the blue and red
sequences of galaxies, affording an exploration of the transition galaxies
between these sequences.  Cosmic downsizing of galaxies hosting active star
formation is also illustrated using {\it GALEX} data, by plotting the
surface density of star formation versus redshift (Figure~18 of
\citet{Martin:07:415}).  These evolution diagnostics are further enhanced
when comparing a large local galaxy sample \citep{Paz:07:185} in the
restframe UV with deep optical high-z galaxy surveys that observe the
younger universe also in the restframe UV.  

What do these diagnostics tell us about galaxies hosting various types of
SNe?  To begin to answer this, we analyze a preliminary set of $\sim100$
galaxies hosting both core-collapse (CC) and thermonuclear (type Ia) SNe.
We derive their integrated UV photometry using a preliminary version of the
{\it GALEX} Large Galaxy Atlas (GLGA, \citep{Seibert:09}).  To this we add
integrated optical and IR photometry extracted from online
databases\footnote{{\tt http://nedwww.ipac.caltech.edu/} and {\tt
http://irsa.ipac.caltech.edu/}}.  We fit the resulting SEDs with star
formation history models \citep{Fioc:97:950, Borgne:02:446, Borgne:04:881},
and plot the photometry on diagnostic diagrams to explore the link between
host evolution and SN production.

\begin{figure}[t]
	\includegraphics[angle=90.,height=0.3\textheight]{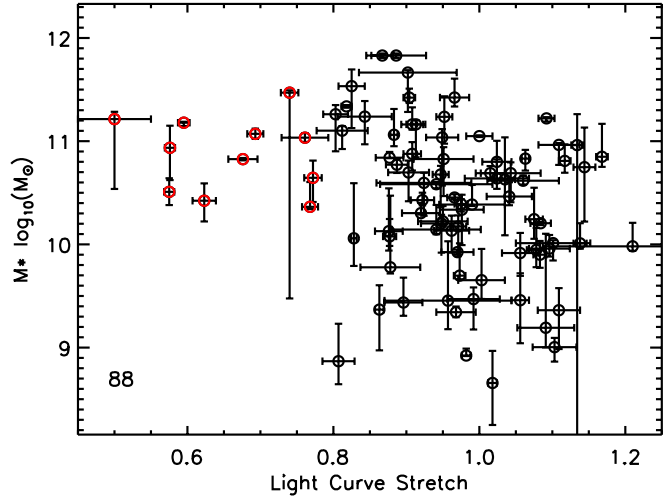}
	\caption{Derived host stellar mass versus SN~Ia light curve stretch for
	the hosts of low-redshift, well-observed SNe~Ia.  Note that for
	hosts of SNe~Ia with stretch $ < 0.8$ (open red circles), the mass 
	range is much narrower than hosts of higher-stretch SNe~Ia (open
	black circles).\label{fig_str_mstar}}
\end{figure}

\section{Host Physical Properties From SED Fits}

We first examine the SED fit results for a set of low-z galaxies hosting
well-observed SNe~Ia and plot the results versus the light-curve stretch
\citep{Conley:08:482}, which parameterizes the evolution timescale of the
SN luminosity.  Figure~\ref{fig_str_mstar} illustrates the trend in host
stellar mass with stretch.  It appears that low-stretch, and thus
low-luminosity, SNe~Ia prefer hosts in a narrow mass range of $10.2 <
\log$~M$_* < 11.5$ (with M$_*$ in M$_{\odot}$), while the higher stretch
SNe~Ia are found in hosts over a much larger range: $8.5 < \log$~M$_* <
12$.

\begin{figure}
	\includegraphics[angle=90.,height=0.3\textheight]{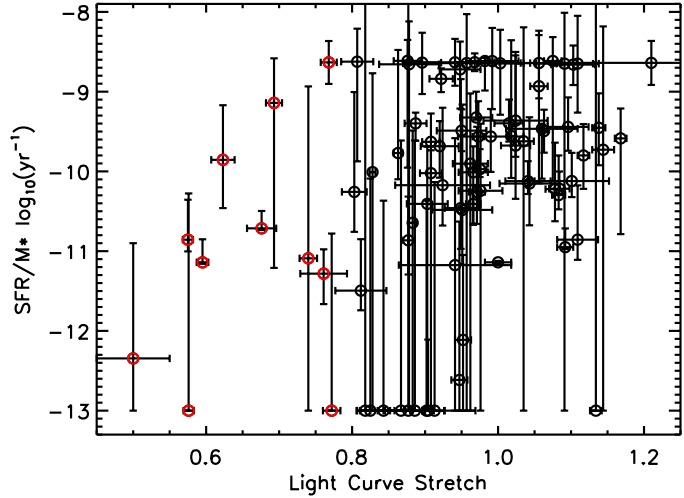}
	\caption{Derived host specific star formation rate (SFR/M$_*$)
	versus stretch for the hosts of low-redshift, well-observed SNe~Ia.
	For hosts of SNe~Ia with $s < 0.8$, there appears to be a downward 
	trend in SFR/M$_*$ as stretch declines.\label{fig_str_ssfr}}
\end{figure}

Figure~\ref{fig_str_ssfr} shows host specific star formation rate (sSFR
$\equiv$ SFR/M$_*$ in yr$^{-1}$) versus stretch.  We find that the hosts of
low-luminosity SNe~Ia have a range of sSFR indicating a range of host
morphology as was found in other studies (e.g., \citep{Howell:01:L193}).
However, here we see indications of a trend in sSFR for hosts of SNe with
stretch $< 0.8$.  These trends in M$_*$ and sSFR are also apparent in much
simpler diagnostic plots such as absolute K-band magnitude and far-UV minus
K-band (FUV-K) color versus stretch, indicating that they are not artifacts
of the SED-fitting process.  The statistical significance of these trends
will be quantified after the photometry and SED-fitting are further refined
\citep{Neill:09}.

\section{The Galaxy Color-Magnitude Diagram}

\begin{figure}
	\includegraphics[angle=90.,height=0.42\textheight]{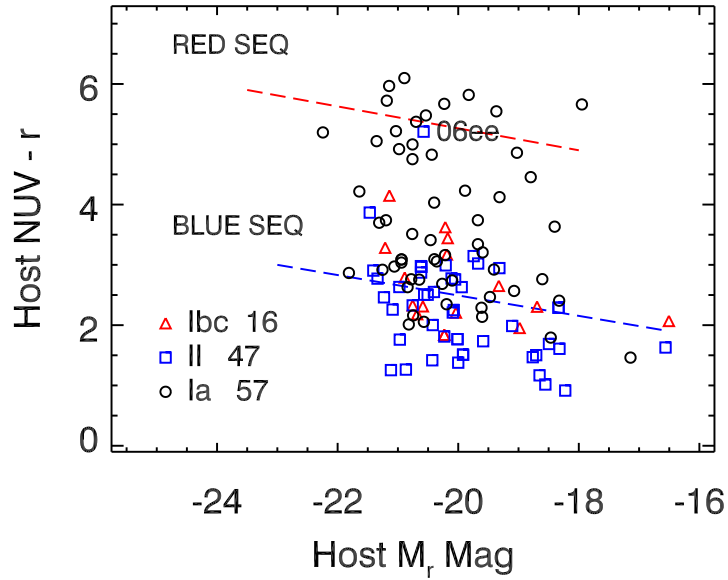}
	\caption{SN host NUV-r color versus absolute r-band magnitude
	(M$_r$) for 130 SN hosts coded by type as follows: type Ia -
	black circles, type II - blue squares, and type Ib/c - red triangles.
	The loci of the red and blue sequences are indicated by the red and
	blue dashed lines.
	\label{fig_wdcmd}}
\end{figure}

We next examine where local SN hosts fall on the galaxy NUV-r versus M$_r$
plane as a function of SN type.  Figure~\ref{fig_wdcmd} plots SN
hosts coded for SN type along with the red and blue galaxy sequences and
should be compared with Figure~9 in \citet{Wyder:07:293}.  

We see that all but one (SN2006ee) of the CC SNe (types II and Ibc) are
found in blue sequence galaxies, while the SNe~Ia appear in galaxies in
both sequences and in between.  In the blue sequence, the type II SN hosts
dominate on the bluest side, while the type Ibc hosts seem to extend into
the `green valley' between the two sequences.  Currently, the sample is too
small to draw strong conclusions about the hosting populations of these
SNe, but the distribution in this diagram appears to contain information
that will be useful in our efforts to constrain progenitors for various
types of SNe.  Once our sample is large enough, we can sub-divide the basic
SN types to see if more structure in this diagram is revealed.

\section{Summary}

We have seen that SED fitting and diagnostic plots derived from integrated
photometry of SN host galaxies have the potential to tell us about the host
evolutionary state and hence constrain the progenitors of SNe of different
types.  We will continue to accumulate high quality SN host photometry so
we can improve the sample size and depth for each host and SN type.  We
will also continue to refine our SED fitting technique and apply this and
galaxy evolution diagnostic techniques to the host photometry.  Our first
efforts will concentrate on the nearby set of SNe~Ia that have well
determined stretch values \citep{Neill:09}, but future efforts will explore
all types of SNe.


\begin{theacknowledgments}
  We acknowledge the use of the NASA Extragalactic Database and the
  Infra-Red Science Archive and their vital contribution to this work.
\end{theacknowledgments}



\bibliographystyle{aipproc}   


\end{document}